# Influence of Cavity Geometry on the Bubble Dynamics of Nucleate Pool Boiling

M.S. Whiting[1,2], W.J. van den Bergh[1], P.E. Theodorakis[3], M. Everts[1,4*]

[1]University of Pretoria, Pretoria, Lynwood, 0028, South Africa
[2]University of Antwerp, Antwerp, Prinsstraat 13, 2000, Belgium
[3]Institute of Physics, Polish Academy of Sciences, Al. Lotników 32/46, 02-668 Warsaw, Poland
[4]University College London, London, WC1E 7JE, United Kingdom
*Corresponding author: m.everts@ucl.ac.uk

## Abstract

Nucleate pool boiling is known for its exceptional heat transfer coefficients, with the use of cavities further improving bubble nucleation and heat transfer rate. To promote this heat transfer enhancement technique, a thorough understanding of the influence of cavity geometry on single bubble dynamics is required. The influence of depth and radius of cylindrical and conical cavities on the bubble dynamics of nucleate pool boiling of R1234yf were numerically investigated. The cavity radius was varied between 50 µm and 400 µm and the cavity depth between 100 µm and 1000 µm at a fixed heat flux of 28 kW/m$^2$. It was found that the maximum equivalent diameter prior to departure was constant for cavities with radii smaller than 120 µm, while it increased linearly when increasing the cavity radius further. Cylindrical cavities exhibited high stability regardless of cavity radius or depth whereas conical cavities showed a decrease in vapour retention with increasing cavity angle. During the necking phase, the bubble interface became pinned at the cavity edge, depending on conical cavity angle, implying that smaller radii allowed for enhanced surface rewetting. Conical cavities could be considered as cylindrical cavities when the cavity angle was less than a quarter of the interface contact angle. When translating the single cavity findings to cavity array design, cylindrical cavities were recommended as they allowed for stable bubble behaviour. For increased nucleation zones and rewetting, a sub-critical radius was recommended. Wider cavities were recommended for high superheat conditions as larger bubbles could enhance bubble growth.

# Nomenclature

**Latin Symbols**

| | | |
|---|---|---|
| $A$ | Area | m² |
| $A_M$ | Constant of Mikic et al. [1] correlation | m/s |
| $B$ | Bubble growth constant | - |
| $B_M$ | Constant of Mikic et al. [1] correlation | m/s$^{1/2}$ |
| $C_D$ | Drag coefficient | - |
| $Co$ | Courant number | - |
| $Cp$ | Specific heat | J/kg·K |
| $D$ | Diameter | mm |
| $d$ | Distance between two points | m |
| $F, \mathbf{F}$ | Force (scalar, vector) | N |
| $f_{Hsu}$ | Function for the Hsu [2] correlation | °C or K |
| $g, \mathbf{g}$ | Gravitational acceleration (scalar, vector) | m/s² |
| $g_c$ | Gravitational correction factor | - |
| $H$ | Cavity depth | µm |
| $h$ | Specific enthalpy | J/kg·K |
| $h_{fg}$ | Specific enthalpy of vaporisation | J/kg·K |
| $Ja$ | Jakob number | - |
| $k$ | Thermal conductivity | W/m·K |
| $L_c$ | Capillary length | m |
| $\dot{m}$ | Mass flow rate | kg/s |
| $\hat{n}$ | Unit normal vector | - |
| $P$ | Pressure | Pa |
| $P_{rgh}$ | Auxiliary pressure | Pa |
| $q$ | Energy density | J/m² |
| $\dot{q}$ | Heat flux | W/m² |
| $R$ | Cavity radius | µm |
| $R_{base}$ | Radius of bubble base | µm |
| $R_{crit}$ | Critical cavity radius | µm |
| $R_{ml}$ | Threshold radius for microlayer criterion | µm |
| $R_{peak}$ | Peak radius of bubble base | µm |
| $S$ | Array spacing | µm |
| $S_r$ | Ratio of microlayer length to slip length | - |
| $T$ | Temperature | °C or K |
| $\Delta T_{sub}$ | Degree of subcooling | °C or K |
| $\Delta T_{sup}$ | Degree of superheating | °C or K |
| $t$ | Time | s |
| $t_g$ | Bubble growth time | s |
| $t^+$ | Scaled time value for Mikic et al. [1] correlation | - |
| $u, \mathbf{u}$ | Velocity (scalar and vector) | m/s |
| $V$ | Volume | m³ |
| $w$ | Weighting function for energy source model | - |
| $x, \mathbf{x}$ | Distance (scalar, vector) | m |
| $y$ | Distance in the vertical direction | m |

**Greek Symbols**

| | | |
|---|---|---|
| $\alpha$ | Volume fraction | - |
| $\dot{\alpha}$ | Rate of change of volume fraction | - |
| $\theta$ | Contact angle | ° |
| $\mu$ | Dynamic viscosity | Pa·s |
| $\nu$ | Kinematic viscosity | m$^2$/s |
| $\pi$ | Mathematical constant | - |
| $\rho$ | Density | kg/m$^3$ |
| $\sigma$ | Surface tension | N/m |
| $\varphi$ | Bubble shape angle for Hsu [2] correlation | ° |
| $\gamma$ | Thermal diffusivity | m$^2$/s |
| $\beta$ | Expansion coefficient | K$^{-1}$ |
| $\delta$ | Dirac delta function | - |

**Subscripts**

| | |
|---|---|
| $A$ | Acceleration |
| $AM$ | Added mass |
| $B$ | Buoyancy |
| $cav$ | In reference to cavity geometry |
| $D$ | Drag |
| $dep$ | Departure |
| $eq$ | Equivalent |
| $eq, dep$ | Equivalent parameter at point of departure |
| $i$ | Placeholder for either liquid or vapour |
| $int$ | Interface |
| $l$ | Liquid |
| $M$ | Momentum |
| $m$ | Counter for weighting function |
| $ml$ | Microlayer |
| $nei$ | Neighbour |
| $P$ | Pressure |
| $pc$ | Phase change |
| $rem$ | Remaining |
| $rw$ | Rewetting |
| $S$ | Surface tension |
| $s$ | Solid |
| $sat$ | Saturated |
| $T$ | Thermal |
| $tbl$ | Thermal boundary layer |
| $tot$ | Total |
| $v$ | Vapour |

**Superscripts**

| | |
|---|---|
| $T$ | Transpose of a matrix |
| $*$ | Non-dimensionalised value |

**Abbreviations**

| | |
|---|---|
| ALE | Arbitrary Lagrangian–Eulerian |
| CFD | Computational fluid dynamics |
| LS | Level set |
| PLIC | Piecewise linear interface calculation |
| RDF | Reconstructed distance functions |
| TBL | Thermal boundary layer |
| VOF | Volume of fluid |

# 1 Introduction

Nucleate pool boiling is a complex liquid-to-vapour phase-change phenomenon that is associated with exceptionally high heat transfer rates at relatively low and stable temperatures. These appealing characteristics have led to nucleate pool boiling being used in a wide array of engineering applications from electricity generation to the cooling of micro-electronics. It has also been the subject of a vast amount of research, with pioneering work done by Nukiyama [3], who developed the first boiling curve for water at atmospheric pressure. Since then, the complexity of nucleate pool boiling has been exhibited as it has been shown to be dependent on many factors including pressure, wall superheat, fluid properties, and surface geometry. Many studies have been devoted to investigating various forms of surface enhancement for the purposes of increasing the heat transfer coefficients. The use of cavities, pin-fins, microchannels, surface roughness, porous structures, porous coatings, and wettability patterns have all been investigated and shown to offer potential improvements to the boiling heat transfer coefficient [4, 5].

Cavities and cavity arrays offer a structured, replicable, and relatively simple way to enhance a boiling surface. They enhance the boiling heat transfer coefficient through several ways, one of which is the promotion of bubble nucleation by means of vapour entrapment. The presence of tiny vapour bubbles/nuclei entrapped by cavities, pores, or scratches on the heated surface encourages bubble nucleation as the nuclei can merge to form a bubble, which is then able to grow. The entrapment potential of conical cavities was first studied by Bankoff [6], who proposed that, for a given wetting fluid with a static contact angle of $\theta$, a conical cavity is capable of sufficiently entrapping vapour such that nucleation is favoured when the conical cavity angle ($\theta_{cav}$ as indicated in Figure 1(b)) is less than half the contact angle, i.e. $\theta_{cav} < \theta/2$. Lorenz [7] came to the same conclusion as Bankoff [6], whereas Tong et al. [8] proposed using the dynamic advancing contact angle as opposed to the static contact angle. Other researchers developed alternative criteria for vapour entrapment inside conical cavities [9, 10], but the review of Mahmoud and Karayiannis [4] highlighted that the Bankoff [6] criterion is sufficiently conservative and appropriate for engineering design.

If a cavity is capable of entrapping vapour (thus satisfying the Bankoff [6] criterion), a bubble may nucleate within this cavity if the wall superheat is sufficient. The required superheat to initiate nucleation has been shown to be related to the cavity radius [2]. Griffith and Wallis [11] did pioneering work in modelling the relationship between cavity size and the required superheat for bubble nucleation. However, their model was limited as it predicted a single active cavity radius for a given wall superheat, whereas Hsu [2] noted that for a given wall superheat, the active nucleation sites have a range of radii. Hsu [2] developed a model to calculate the range of active cavity radii ($R$) based on the surface superheat ($\Delta T_{sup}$), liquid subcooling ($\Delta T_{sub}$), and thermal boundary layer thickness ($d_{tbl}$):

$$R_{(max,min)} = \frac{d_{tbl}\sin(\varphi)\Delta T_{sup}}{2f_{Hsu}}\left[1 \pm \sqrt{1 - \frac{8\sigma T_{sat}f_{Hsu}}{d_T \rho_v h_{fg}\Delta T_{sup}^2}}\right] \quad (1)$$

$$f_{Hsu} = (1 + \cos(\varphi))(\Delta T_{sup} + \Delta T_{sub}) \quad (2)$$

$$\varphi = \theta + 90° - \theta_{cav} \quad (3)$$

where $\rho_v$ is the vapour density, $h_{fg}$ is the enthalpy of vaporisation, $T_{sat}$ is the saturation temperature, $\sigma$ is the interface surface tension, $\theta$ is the interface contact angle and $\theta_{cav}$ is the cavity angle (taken to be zero for cylindrical cavities).

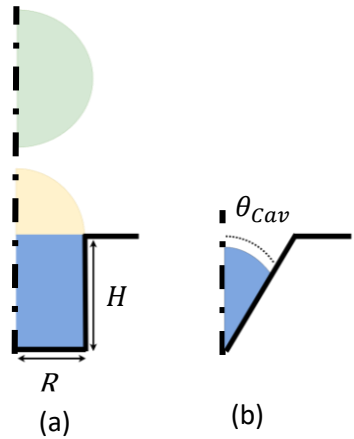

*Figure 1: Schematic of (a) cylindrical and (b) conical cavity geometries indicating cavity depth, radius, and angle. The blue area represents vapour within the cavity, yellow area represents the bubble volume that has emerged from the cavity but not detached, and the green area represents a detached bubble.*

Shoji and Takagi [12] experimentally investigated the effect of conical, cylindrical, and re-entrant cavities created on a copper plate on the single bubble dynamics of saturated water. They found that both cylindrical and re-entrant cavities exhibited higher stability. Conical cavities were prone to periods of inactivity and required high superheat to maintain bubbling, although it should be noted that their conical cavities had large cavity angles of approximately 59° and 63.5°.

Narayan et al. [13] investigated the effect of cylindrical cavity depth on the bubble dynamics of saturated water on an aluminium plate. Although the cavity depth had a negligible impact on the bubble departure diameter, it affected the departure frequency with shallower cavities leading to earlier departure. The departure frequency was shown to be consistent for deeper cavities, but as the level of superheat increased, variation in departure frequency increased with shallower cavities resulting in significantly more deviation. In addition, for deep cavities, the maximum bubble base diameter was independent of superheat, whereas the more rapid and chaotic growth from shallower cavities resulted in the maximum bubble base diameter varying with surface superheat.

The impact of conical cavity angle has also recently been investigated by Narayan et al. [14] for conical cavities with a constant depth. Comparing with the results from a cylindrical cavity with similar dimensions, it was shown that bubble departure diameter was similar for cavities with similar diameters, regardless of their shape. It was also shown that conical cavities with larger cavity angles resulted in longer bubble growth periods.

Apart from the work done by the groups of Shoji et al. [12, 15, 16] and Narayan et al. [13, 14], little literature could be found detailing the influence of cavity geometry on the bubble dynamics of single bubble nucleate pool boiling. While these studies have provided insight into the influence of cavity geometry, the range and resolution of cavity depths and radii investigated have been limited. Although it is experimentally challenging to manufacture and test a wide array of cavity sizes, it can be more efficiently investigated numerically.

In recent years, the volume of fluid (VOF) method has seen much use in the numerical study of nucleate pool boiling. Influential work was done by the group of Stephen and Kunkelmann [17-19], who developed a sub-model to account for microlayer evaporation [20]. They implemented the phase change model developed by Hardt and Wondra [21] which was developed for both VOF and level set (LS) applications and is now a standard model in phase change literature. Sato and Ničeno [22] utilised an equivalent method to VOF that they refer to as a "colour function" to develop a depletable microlayer model [23] and simulate nucleate pool boiling from the single bubble regime at low heat fluxes, up to the "vapour mushroom" regime at higher heat fluxes [24]. The model of Sato and Ničeno [22] has shown good agreement with experimental data and the capability to model the transition between boiling regimes [25].

The VOF method is capable of accurately modelling nucleate pool boiling under a variety of conditions while including evaporation from the microlayer and conjugate heat transfer. Despite its pervasive use, few studies could be found which utilised the VOF method to investigate the influence of surface geometry on nucleate pool boiling. With limited information regarding the fundamentals of how the cavity geometry impacts bubble dynamics, the design and optimisation of boiling surfaces becomes difficult. Furthermore, as will be shown in this paper, existing correlations cannot accurately predict the bubble dynamics from different cavity geometries. Hence, the purpose of this study was to numerically investigate the influence of cylindrical and conical cavity geometries on the bubble dynamics of a single bubble under nucleate pool boiling conditions. A wide array of cavity radii, depths and angles were investigated to identify critical cavity radii as well suggestions to optimise rewetting heat transfer.

## 2 Numerical Modelling

### 2.1 Governing Equations

In this study, all simulations were performed using the OpenFOAM v2106 framework. The *multiRegionPhaseChangeFlow* solver, which forms part of the TwoPhaseFlow libraries created by Scheufler and Roenby [26] was utilised. The libraries make use of the Volume of Fluid (VOF) capturing method with the *multiRegionPhaseChangeFlow* solver taking compressibility, phase change, surface tension, and conjugate heat transfer into account.

While the solver could account for compressibility, the maximum velocity in this study was a fraction of a meter per second corresponding to a Mach number in the order of $10^{-3}$, which is low enough to consider the system as incompressible. With this assumption, the phase continuity equation was given as follows:

$$\frac{\partial \alpha}{\partial t} + \nabla \cdot (\boldsymbol{u}\alpha) = \dot{\alpha}_{pc} \qquad (4)$$

with $\alpha$ being the volume fraction, $\boldsymbol{u}$ being the velocity field and $\dot{\alpha}_{pc}$ the source term accounting for phase change.

For laminar flow, the Navier–Stokes momentum equation is reduced to:

$$\frac{\partial (\rho \boldsymbol{u})}{\partial t} + \nabla \cdot (\rho \boldsymbol{u}\boldsymbol{u}) - \nabla \cdot \{\mu(\nabla \boldsymbol{u} + (\nabla \boldsymbol{u})^T)\} = -\nabla P_{rgh} + (\boldsymbol{g} \cdot \boldsymbol{x})(\rho_l - \rho_v)\hat{\boldsymbol{n}}_{int}\delta_{int} + \boldsymbol{F} \qquad (5)$$

where $\rho$ was the mass density, $\mu$ the dynamic viscosity, $\boldsymbol{g}$ the gravitational direction vector (with $\boldsymbol{x}$ being the relative position vector), $\hat{\boldsymbol{n}}_{int}$ the unit vector normal to the multiphase interface which points into the heavier fluid, and $\delta_{int}$ the Dirac delta function. The vector $\boldsymbol{F}$ accounted for additional force source terms such as surface tension. The *l* and *v* subscripts indicated the liquid and vapour phase, respectively. The $P_{rgh}$ term was the pressure, excluding hydrostatic effects, commonly used in buoyant and multiphase cases:

$$P_{rgh} = P - (\boldsymbol{g} \cdot \boldsymbol{x})\rho \qquad (6)$$

with *P* being the pressure. This pressure formulation was numerically convenient, allowing for simpler boundary condition specification and for the remaining gravity term containing the Dirac delta to be treated as zero at all points not at the interface.

The energy equation was written in terms of temperature, $T_i$, where the subscript *i* denoted either *l* or *v* for the liquid and vapour formulations, respectively:

$$\frac{\partial \alpha_i \rho_i Cp_i T_i}{\partial t} + \nabla \cdot (\alpha_i \rho_i Cp_i T_i \boldsymbol{u}) = \nabla \cdot (k_i \nabla T_i) + \dot{q}_{pc,i} \quad (7)$$

where *Cp* denoted the specific heat, *k* the thermal conductivity, and $\dot{q}_{pc}$ a source term representing the change in energy due to phase change.

The model accounted for conjugate heating with the heat transfer in the solid region written in terms of specific enthalpy, $h_s$, with the subscript *s* indicating the solid region:

$$\frac{\partial \rho_s h_s}{\partial t} - \nabla \cdot \left(\frac{k_s}{Cp_s} \nabla h_s\right) = 0 \quad (8)$$

The solver accounted for phase change in two discrete steps. Firstly, the energy source term associated with phase change, $\dot{q}_{pc}$ (from Equation 7), was calculated using an implicit gradient-based model based on the model by Batzdorf [27]. The model assumed that the interface was at saturation temperature and considered the heat transfer from several neighbouring cells. The energy source term was calculated separately for each phase, using Equation 9, with the positive interface normal vector used to take the liquid cell contributions into account and the negative interface normal vector used for the vapour:

$$\dot{q}_{pc,i} = \sum_{nei} \frac{w k_i}{d_{nei}} T_{sat} - \sum_{nei} \frac{w k_i}{d_{nei}} T_{nei} \quad (9)$$

with $T_{nei}$ being the temperature of a neighbouring cell, $d_{nei}$, being the minimum distance between the interface segment and the neighbouring cell centre and *w* being a weighting function:

$$w = \left(\frac{\cos \theta_{nei}}{\sum_m \cos \theta_m}\right)^4 \text{ with } \theta_{nei} = \hat{\boldsymbol{n}}_{int} \cdot (\boldsymbol{x}_{nei} - \boldsymbol{x}_{int}) \quad (10)$$

The denominator of the weighting function was a sum over all *m* neighbouring cells with $\theta_m$ being calculated in the same way as $\theta_{nei}$. After calculating the energy source term, the mass source term ($\dot{m}$) was calculated at the interface and smeared over the neighbouring interface cells using a modified form of the Hardt and Wondra [21] model:

$$\dot{m} = \frac{\dot{q}_{pc}}{h_{fg}} \quad (11)$$

with $h_{fg}$ being the enthalpy of vaporisation.

The interface curvature was calculated using the Reconstructed Distance Function (RDF) model [28], and the interface normal was found from the phase fraction field using the Continuum Surface Force (CSF) method of Brackbill et al. [29]:

$$\hat{\boldsymbol{n}}_{int} \delta_{int} = \nabla \alpha \quad (12)$$

The continuity equation was solved using the isoAdvector algorithm, a geometric VOF solver, developed by Scheufler and Roenby for OpenFOAM [30], with the plicRDF method being used for the interface reconstruction [31]. The PIMPLE algorithm was used for solving the pressure–velocity coupling. A variable time-step was used, and the length of each time-step was determined such that the maximum Courant number, Co, did not exceed 0.1. The Courant number was defined as:

$$Co = \frac{u \Delta t}{\Delta x} \tag{13}$$

with $\Delta t$ being the time-step duration, $u$ the magnitude of the flow velocity, and $\Delta x$ the cell length in the direction of flow.

## 2.2 Boundary Conditions and Initial Conditions

A schematic of the calculation domain showing the boundary conditions is given in Figure 2. The domain was axisymmetric, implying the front and back surfaces were treated as symmetry boundary conditions, while the left-hand surface was the axis. Separate solid and fluid domains allowed for the modelling of conjugate heat transfer. The bottom surface was heated with a constant heat flux while a zero-gradient temperature boundary condition was used for the top and right-hand boundaries. The fluid–solid interface had a no-slip velocity, a fixed flux pressure, and a constant contact-angle of 54°. Following Kunkelmann [19], the fluid domain's right-hand boundary was treated as a slip-wall, while the top boundary assumed constant pressure. To account for conjugate heating, the mesh was split into a fluid and a solid region, with the upper-horizontal surface between the two regions situated at $y = 0$, implying the cavity cells had negative $y$-values.

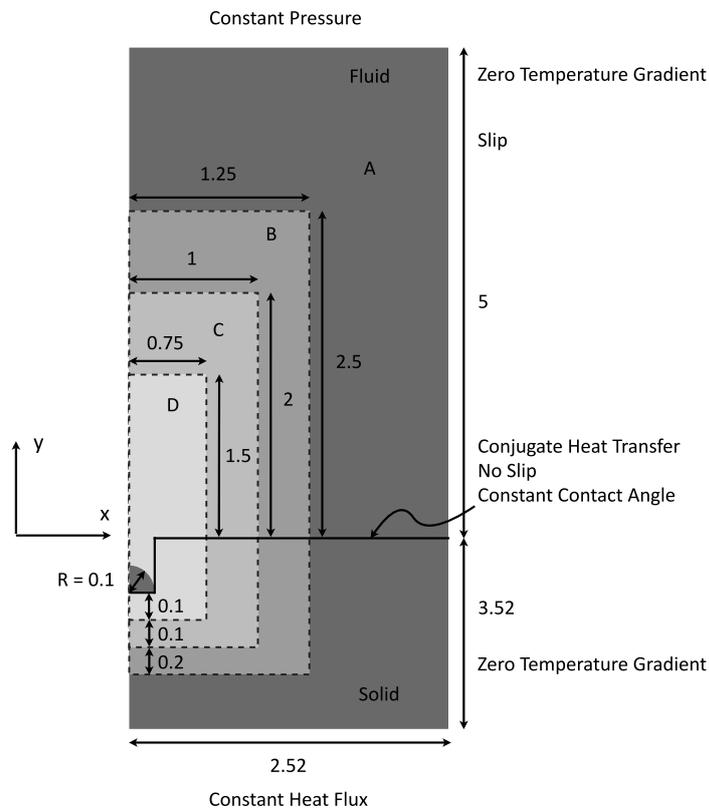

Figure 2: Mesh domain with dimensions (in mm) and boundary conditions indicated. The mesh is divided into four zones to allow for variable mesh refinement with zone A having a cell size of 40 μm, zone B: 20 μm, zone C: 10 μm and zone D: 5 μm.

When it came to initialising the fluid temperature field, a common method, which is to prescribe a thermal boundary layer where the temperature linearly decreases from the wall temperature, $T_{wall}$, to the fluid's saturation temperature, $T_{sat}$. The thickness of the thermal boundary layer is commonly determined using the correlation of Kays and Crawford [32]:

$$d_{tbl} = 7.14 \left(\frac{\nu_l \gamma_l}{g \beta_T \Delta T}\right)^{1/3} \quad (34)$$

with $\nu_l$ and $\gamma_l$ representing the liquid's kinematic viscosity and thermal diffusivity, respectively, and $\beta_T$ being the coefficient of thermal expansion. While use of this correlation is pervasive (see the work of Son et al. [33] for an example), it is primarily applicable to simulating nucleate pool boiling on a flat surface. When investigating cavity geometries, Equation 14 became challenging to use as the thermal boundary layer must be applied in different directions and blended around corners to ensure a smooth fluid temperature field.

Therefore, a single-phase simulation with liquid only (i.e., prior to seeding the bubble) was first conducted to obtain a realistic temperature field and investigate the influence of cavity geometry on the initial thermal boundary layer. When using this method, the single-phase simulation's resulting thermal boundary layer could be controlled by varying the initial temperature of the solid region and the duration of the single-phase simulation, which allowed the initial conditions to be fine-tuned to match experimental results. Thereafter, a hemi-spherical bubble with a radius of 100 µm was seeded at the centre of the cavity's bottom surface and the two-phase simulation begun. Similar bubble seed sizes have also been used in the literature [19, 34]. The volume of the bubble seed was at most 0.14% of the bubble volume at departure, therefore it was sufficiently small to have a negligible impact on the final departure mechanics.

## 2.3 Cavity Geometry and Investigation Parameters

Cylindrical and conical cavities were illustrated in Figure 1, showing the relevant nomenclature to the cavity geometry, while The cavity sizes and various departure metrics have been presented in both dimensional and non-dimensional (indicated using *) forms using the capillary length ($L_c$), as defined in Equation 17, which is equal to 0.773 mm for R1234yf. The capillary length was used as it is often linked with the departure diameter of a bubble (most departure diameter correlations reviewed in Mahmoud and Karayiannis [4] include the capillary length, for example) and has been shown to be a useful consideration in the design of boiling surfaces [35].

$$L_c = \sqrt{\frac{\sigma}{\Delta \rho g}} \quad (4)$$

Table I summarises the radii and depths that were investigated. The conical cavity angle was not explicitly chosen but resulted from the combination of cavity radius and depth. This opportunity is also taken to clarify the nomenclature going forward and how the results were processed. The bubble equivalent diameter is denoted $D_{eq}$, and represents the diameter of a sphere with the same volume ($V$) as the bubble:

$$D_{eq} = \left(\frac{6V}{\pi}\right)^{1/3} \quad (15)$$

For this study the bubble equivalent diameter only considers the vapour volume that had emerged from the cavity, i.e., in the positive-$y$ portion of the domain, including both the yellow and green vapour volumes represented in Figure 1(a). When investigating the bubble growth within the cavity, the entire domain (including the negative-$y$ portion of the domain, i.e., the blue vapour region in Figure 1) is considered, resulting in the "total equivalent diameter", $D_{tot}$. When investigating the bubble departure, literature generally refers to the bubble "departure diameter" as the equivalent diameter of the bubble that detaches from the surface. The same definition was used here, and it is denoted with $D_{dep}$, represented with the green vapour volume in Figure 1(a). However, as not all the bubble volume detaches, there is a discrepancy between the "departure diameter" and the "equivalent diameter at departure" because the equivalent diameter includes the entirety of the positive-$y$ domain (both green and yellow). To avoid confusion, the symbol $D_{eq,dep}$ is used to signify the equivalent diameter of the vapour volume in the positive-$y$ domain at the time of departure. The bubble that remains attached to the surface, $D_{rem}$, is then used in combination with $D_{eq,dep}$ to determine the departure diameter:

$$D_{dep} = D_{eq,dep} - D_{rem} \tag{16}$$

The cavity sizes and various departure metrics have been presented in both dimensional and non-dimensional (indicated using *) forms using the capillary length ($L_c$), as defined in Equation 17, which is equal to 0.773 mm for R1234yf. The capillary length was used as it is often linked with the departure diameter of a bubble (most departure diameter correlations reviewed in Mahmoud and Karayiannis [4] include the capillary length, for example) and has been shown to be a useful consideration in the design of boiling surfaces [35].

$$L_c = \sqrt{\frac{\sigma}{\Delta \rho g}} \tag{4}$$

Table I: Cavity size test matrix detailing the cavity depths, cavity radii, and resulting conical cavity angles tested. A fixed input heat flux of 28 kW/m² was used.

| Cavity Radius, $R$ [μm] | Cavity Depth, $H$ [μm] | Cavity Radius, $R^*$ | Cavity Depth, $H^*$ | Conical Cavity Angle, $\theta_{cav}$ [°] |
|---|---|---|---|---|
| 50 | 250 | 0.065 | 0.324 | 11.3 |
|  | 500 |  | 0.647 | 5.7 |
|  | 1000 |  | 1.294 | 2.9 |
| 100 | 100 | 0.129 | 0.129 | 45.0 |
|  | 250 |  | 0.324 | 21.8 |
|  | 500 |  | 0.647 | 11.3 |
|  | 1000 |  | 1.294 | 5.7 |
| 200 | 100 + | 0.259 | 0.129 | 63.4 |
|  | 250 |  | 0.324 | 38.7 |
|  | 500 |  | 0.647 | 21.8 |
|  | 1000 |  | 1.294 | 11.3 |
| 250 | 100 + | 0.324 | 0.129 | 68.2 |
|  | 250 |  | 0.324 | 45.0 |
|  | 500 |  | 0.647 | 26.6 |
|  | 1000 |  | 1.294 | 14.0 |
| 400 | 100 + | 0.518 | 0.129 | 76.0 |
|  | 250 |  | 0.324 | 58.0 |
|  | 500 |  | 0.647 | 38.7 |
|  | 1000 |  | 1.294 | 21.8 |

+ These cavities were only tested for conical cases to investigate the effect of large cavity angles.

While the chosen cavity dimensions are comparable to previous studies [13, 14, 36], the large cavity angles of some conical cavities could make nucleation unlikely, therefore the Hsu model (Equation 1) was used to verify that nucleation can be expected. For the cylindrical cavities at a superheat of 6.26 K (taken from validation case [36]), the Hsu model predicted nucleation in cavities with radii ranging from 0.06 µm to 651.22 µm, implying nucleation should be possible for all cylindrical cavities considered in this study. For the conical cavity with a cavity angle of 76° (the worst-case scenario), the model predicted nucleation in cavities with radii ranging from 0.10 µm to 142.64 µm, implying that nucleation is unlikely in some of the conical cavities. However, the cavities with radii falling outside the range of the Hsu model remained part of the test matrix with the rationale being that modern surface treatments can be utilised to create sub-cavities or nanopores that may provide nucleation points within the cavities, similar to the work of Deng et al. [37]. Explorations into how nanopore nucleation may be incorporated in micro-cavities may be achieved using Molecular Dynamics [38, 39].

## 2.4 Calculation Domain and Mesh Independence

The domain and mesh formulation presented in Figure 2 contained four zones to allow for variable mesh refinement. The mesh was 2D axisymmetric, encompassing a 5° segment centred at the $y$-axis. The impact of the domain size on the bubble departure diameter and departure time was investigated using the departure metrics associated with the smallest fluid domain as a reference. When comparing the results in Table II, it follows that the bubble departure metrics were negligibly impacted by the domain size, all being within 1% of the reference. Therefore, similar to the study of Li et al. [34], the fluid domain with a height of 5 mm and a width of 2.52 mm was selected. The solid domain had a height of 3.52 mm which had a negligible influence on the heat transfer in this system.

*Table II: Comparison of bubble departure diameter and time using different domain radii and domain heights. The departure metrics associated with the smallest fluid domain was used as a reference.*

| Domain Radius [mm] | Domain Height [mm] | $\Delta D_{eq, dep}$ [%] | $\Delta t_{dep}$ [%] |
|---|---|---|---|
| 1.28 | 2.52 | - | - |
| 1.28 | 5 | 0.03 | 0.28 |
| 1.28 | 10 | 0.02 | 0.85 |
| 2.52 | 2.52 | 0.06 | 0.39 |
| 5.00 | 2.52 | 0.14 | 0.22 |

For the cylindrical cavities, the meshing was straightforward, using a structured hex-mesh (Figure 3(a)) and only relying on unstructured mesh cells at the interfaces between the zones of different cell sizes. For the conical cavities, the snappyHexMesh utility of OpenFOAM was used to capture the diagonal line of the cavity surface, introducing triangular cells along this line (Figure 3(b)). For the conical cavities, a sharp point at the apex of the cone created highly skewed, non-orthogonal cells which resulted in numerical instability. It was considered reasonable to flatten this apex as many fabrication techniques would fail at rendering a perfectly pointed conical cavity.

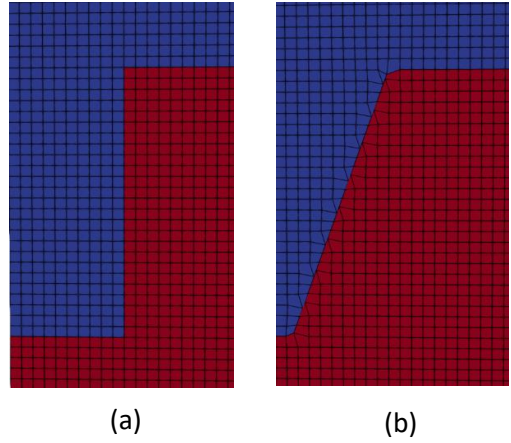

*Figure 3: Sample images contrasting the structure of the mesh generated for (a) cylindrical cavity and (b) conical cavity.*

A mesh independence study was performed investigating the impact of cell size on the departure dynamics. As the mesh was divided into four zones, it allowed for variable refinement, with fine cells in the region of bubble growth and progressively coarser cells further away. This was done to reduce the total cell count and obtain a faster solution time. The coarsest mesh size was 50 μm which was implemented throughout the calculation domain. For smaller cell sizes, zone A (Figure 2), had a cell size of 40 μm, which was then halved to 20 μm in zone B, 10 μm in zone C, and finally 5 μm in zone D. Since the cavity size considered in the mesh independence study had a radius of 100 μm, 40 μm cells could not accurately capture the cavity's geometry resulting in only cell sizes of 5 μm, 10 μm, 20 μm and 50 μm having been tested.

Three metrics were used to compare the results: (1) equivalent bubble diameter at departure ($D_{eq,\,dep}$), (2) bubble growth time ($t_{dep}$), and (3) equivalent diameter of the remaining vapour attached to the surface after departure ($D_{rem}$). Table III summarises the results of the mesh independence study, taking the finest (5 μm) mesh as a reference. Figure 4 visualises these results to show the difference in growth rates. As the results from the 10 μm mesh closely agreed with those of the 5 μm mesh, with all departure metrics being within 5% for the two meshes, it was concluded that acceptable mesh independence was achieved using a 10 μm cell size. This also corresponded well with cell sizes used in similar studies in literature [19, 24, 34].

*Table III: Results of mesh independence study using different mesh cell sizes.*

| Refinement Level [μm] | Δ$D_{eq,\,deq}$ [%] | Δ$t_{dep}$ [%] | Δ$D_{rem}$ [%] |
|---|---|---|---|
| 5 | - | - | - |
| 10 | 0.93 | 1.24 | 4.09 |
| 20 | 2.40 | 14.99 | 7.35 |
| 50 | 3.54 | 40.73 | 44.58 |

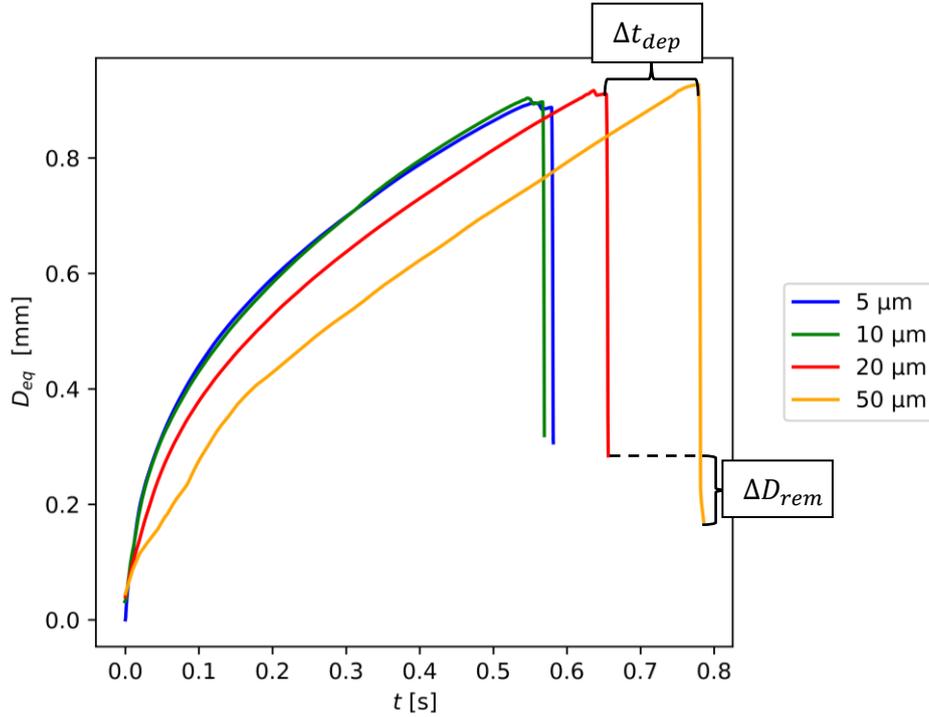

*Figure 4: Bubble growth curves showing the results of the mesh independence study for minimum cell sizes of 5 μm (blue), 10 μm (green), 20 μm (red), 50 μm (orange). Also, shown is how the difference in growth time and remaining bubble diameter were measured.*

## 2.5 Validation of the Numerical Model

The experimental results of Li et al. [36], using R1234yf with an input heat flux of 28 kW/m$^2$, were used for the validation. To achieve single bubble nucleation, Li et al. [36] made use of a cylindrical cavity with a depth of 500 μm and a radius of 100 μm, drilled into the centre of the top surface of a 5 mm-diameter copper rod.

The initial single-phase simulations had an arbitrarily selected runtime of 0.24 s, and various initial solid superheats were considered. Thereafter, the results from the two-phase simulations were compared with the validation case and the initial conditions were fine-tuned to achieve agreement with the experimental data. The validation results of numerical model and initial conditions are summarised in Table IV and visualised in Figure 5. The experimental data [36] and $D_{rem}$ values of the numerical simulations show an initial bubble diameter of 0.31 mm due to a small portion of vapour remaining attached to the cavity after the previous bubble departure. Due to limits in computational power and the possibility of complete vapour detachment from certain cavities, the numerical results presented in this study encompass only the first bubble's lifecycle. This creates a discrepancy in the initial bubble size of the experimental and numerical results. As indicated in Figure 5, the discrepancy in the initial bubble size is overcome by considering numerical results after the point that they reached the initial experimental bubble size. The equivalent diameter at departure and the bubble remaining after departure were in close agreement with the experimental results (within 4%) regardless of the initial temperature of the solid. The bubble size after departure was compared with the first datum of the experimental results, as this would have been recorded after the departure of the previous bubble. The only metric significantly influenced by the initial temperature of the solid was the bubble growth time. An initial surface superheat of 4.85 K was shown to yield the best overall results with all departure metrics being within 2% of the experimental results.

*Table IV: Results of validation study using different initial temperatures for the solid domain.*

| Initial Superheat of Solid Region [K] | $\Delta D_{eq,\ deq}$ [%] | $\Delta t_{dep}$ [%] | $\Delta D_{rem}$ [%] |
|---|---|---|---|
| 3.28 | 2.17 | 20.63 | 3.72 |
| 4.28 | 1.80 | 5.41 | 3.13 |
| 4.85 | 1.69 | 0.71 | 1.86 |
| 5.28 | 1.40 | 10.81 | 0.70 |

While the departure metrics of the numerical results were within 2% of the experimental results, it follows from Figure 5 that the equivalent diameter had large discrepancies (up to 18%) in the early stages of the bubble lifecycle. A possible reason for this disparity could be the negligence of the microlayer heat transfer. Li et al. [34] performed a numerical investigation and achieved very close agreement to their experimental results [36] when implementing the microlayer model of Chen and Utaka [40]. However, due to the relatively slow growth of R1234yf bubbles at the given superheat, the criterion developed by Bures and Sato [41] was used to verify whether a microlayer was likely to develop after a bubble emerged from a cavity. The criterion states that the microlayer contracts (leading to microlayer destruction) for bubbles with radii larger than the threshold ($R_{ml}$) stated in Equation 18.

$$R_{ml} \leq 18B^2 \ln(S_r) \frac{\mu_l \gamma_l}{\sigma} \frac{\sin(\theta_{ml})}{\theta_{ml}^3} \qquad (18)$$

where *B* is a growth factor of 2.096 (for the current case) determined using the method of Scriven [42]. $S_r$ is the ratio of the length of the microlayer to the slip length, and was found to be approximately 7.2 in the investigation of Bures and Sato [41], who also noted it as being difficult to determine accurately. $\theta_{ml}$ is the contact angle of the microlayer. For the current case, even a conservative estimate of $S_r$ = 1000, yielded a threshold radius, $R_{ml}$, of 0.52 μm. As this is significantly smaller than the cavity radius, it could be assumed that no microlayer developed for bubbles emerging from the cavity. Although the possibility of microlayers developing within the cavity existed for a newly nucleated bubble, to the authors' best knowledge, there is no literature showing how microlayers develop or influence bubble growth within cavities. It was therefore assumed that the microlayer growth within the cavity did not significantly affect the final bubble departure (which was the main focus of this paper). The variations in the initial growth rates were assumed to be due to the differences in initial conditions.

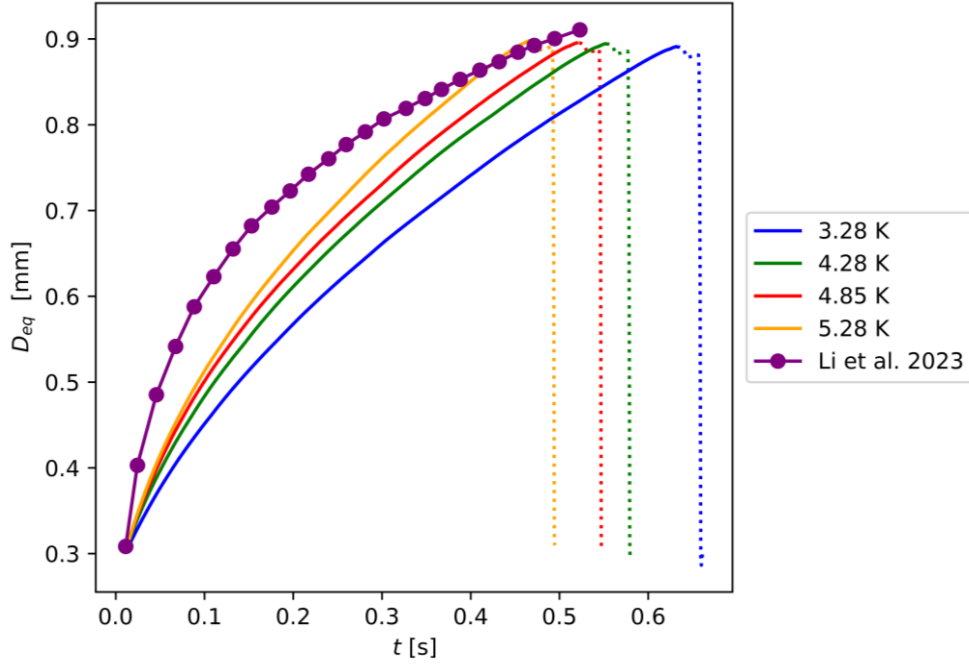

*Figure 5: Bubble growth curves showing the results of the validation and initial condition study featuring the experimental results of Li et al. [36]. The solid lines represent bubble growth until departure and dotted lines continue until bubble leaves the domain.*

## 3   Results: Transient Bubble Growth and Dynamics

### 3.1   Cavity Radius

The transient lifecycle of a bubble within a cylindrical cavity was investigated for different cavity depth and cavity radius combinations, as summarised in The cavity sizes and various departure metrics have been presented in both dimensional and non-dimensional (indicated using *) forms using the capillary length ($L_c$), as defined in Equation 17, which is equal to 0.773 mm for R1234yf. The capillary length was used as it is often linked with the departure diameter of a bubble (most departure diameter correlations reviewed in Mahmoud and Karayiannis [4] include the capillary length, for example) and has been shown to be a useful consideration in the design of boiling surfaces [35].

$$L_c = \sqrt{\frac{\sigma}{\Delta \rho g}} \tag{4}$$

Table I. The results are summarised in Figure 6 using different colours for different radii and different line types to distinguish between different depths.

Two general growth patterns were recognised when comparing the curves depicting the effect of cavity radius on $D_{tot}$ with time. The first growth pattern occurred for all cavities with a radius up to and including 100 µm, from now on referred to as "narrow cavities." These curves were closely grouped and showed a consistent bubble size at departure. The second growth pattern occurred for cavities with a radius larger than 100 µm, from now on referred to as "wide cavities". The curves for the wide cavities had a different trend to those of the narrow cavities. The differences in the initial growth rates of the narrow and wide cavities were due to the bubble initialization radius of 100 µm. This implied that for narrow cavities, the bubble was already in contact with the cavity wall at initialization. Conversely, for wide cavities, the bubble was surrounded by superheated liquid, typically resulting in an increased growth rate when compared with the narrow cavities. This variation in growth rate was not observed for the bubble growth within conical cavities, as the bubble remained in contact with the cavity wall regardless of the cavity radius. These artificial differences in the initial bubble growth rates do not account for the variation in bubble departure size observed for bubbles emerging from wide cavities, nor do they significantly impact the bubble growth after emergence from the cavity.

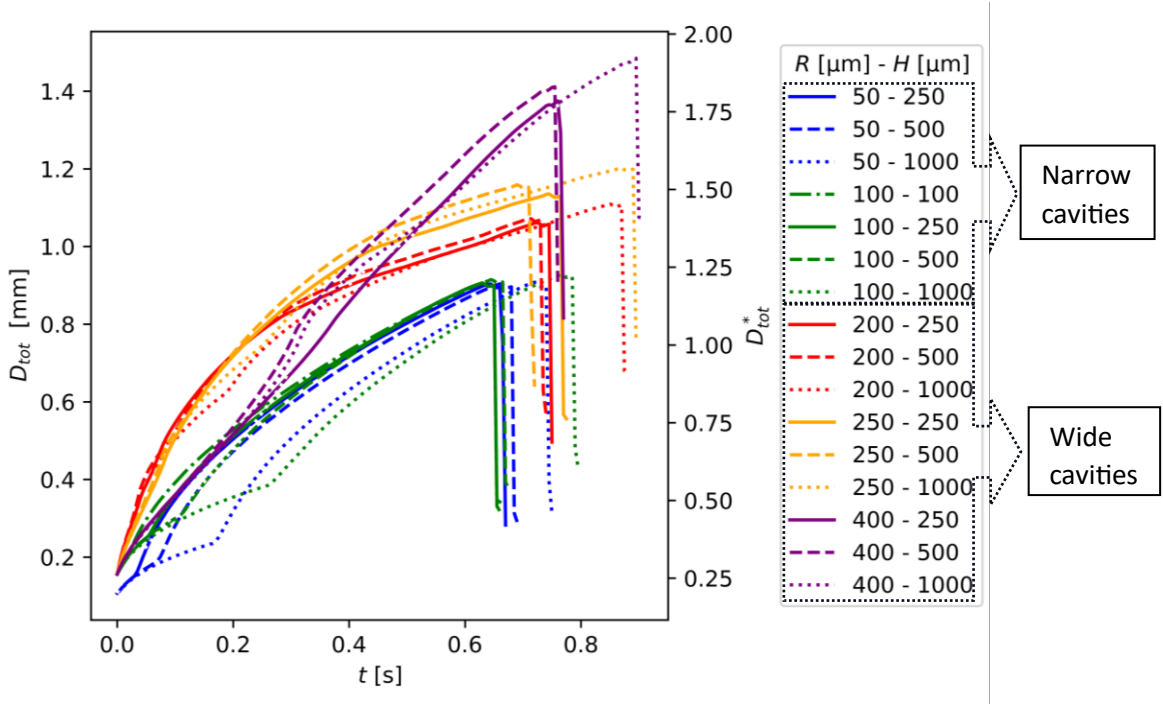

*Figure 6: Bubble growth curves of the total equivalent diameter for cylindrical cavities with solid lines representing a cavity depth of 250 μm, dashed: 500 μm, dotted: 1000 μm and dash-dot: 100 μm. Line colours indicate cavity radius with blue: 50 μm, green: 100 μm, red: 200 μm, orange: 250 μm, purple: 400 μm.*

Selected cavities of uniform depth, compared with existing correlations (summarised in Table V) in Figure 7, show the effect of cavity radius on the departure diameter for both cylindrical and conical cavities. Figure 7(a) shows that for narrow cylindrical cavities (radius of 50 μm and 100 μm), there was negligible difference in their respective departure diameters. Once a critical cavity radius was exceeded, larger bubbles were attainable with an increase in cavity radius. The same trend is observed in Figure 7(b) for bubbles departing from conical cavities, although for larger cavity angles the trend deviates with bubbles being smaller at departure. The three-dimensional summary of the effect of cavity geometry on the departure diameter in Figure 8 better represents this discontinuous behaviour. The narrow cavities had a constant departure diameter of about 900 μm. Once the cavity radius increased beyond 100 μm, the departure diameter increased linearly with the cavity radius. The piecewise relationship between the departure diameter and cavity radius is defined in Equation 22 with all units being in μm.

*Table V: Correlations for the prediction of single bubble growth in nucleate pool boiling.*

| Author | Correlation | |
|---|---|---|
| Sakashita [43] | $D_{eq} = 4.898 Ja^{1/2}\sqrt{\gamma_l t}$ | (19) |
| Cole & Shulman [44] | $D_{eq} = 5 Ja^{3/4}\sqrt{\gamma_l t}$ | (20) |
| Mikic et al. [1] | $D_{eq} = \dfrac{4 B_m^{\,2}}{3 A_M}[(t^+ + 1)^{1.5} - (t^+)^{1.5} - 1]$ | (21) |

| | $A_M = \sqrt{\dfrac{2\rho_v h_{fg}\Delta T_{sup}}{3\rho_l T_{sat}}}$; $B_m = \sqrt{\dfrac{12\gamma_l}{\pi}}Ja$; $t^+ = \dfrac{tA_M{}^2}{B_m{}^2}$ | |

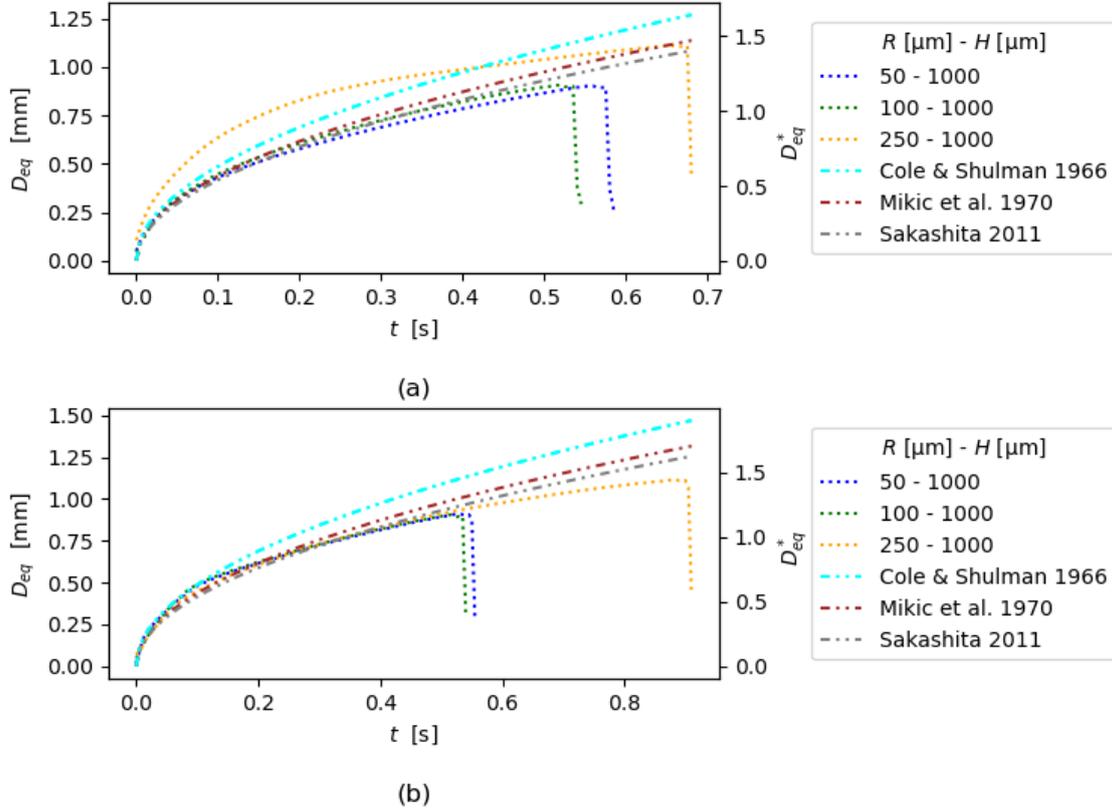

Figure 7: Bubble growth curves of the equivalent diameter highlighting the influence of (a) cylindrical and (b) conical cavity radius on bubble growth and departure for cavities with radii of 50 μm (blue), 100 μm (green) and 250 μm (orange). All cavities are 1000 μm deep.

The discontinuity between the behaviour of bubbles growing from narrow and wide cavities is quantified in absolute (Equation 22) and non-dimensionalised (Equation. 23) form with the interesting characteristic of the constant in the linear portion of the equation being nearly equal to the capillary length.

$$D_{eq,dep} = \begin{Bmatrix} 900 & if\ R < R_{crit} \\ 1.373R + 765 & if\ R > R_{crit} \end{Bmatrix} \qquad (22)$$

$$D_{eq,dep} \approx \begin{Bmatrix} 1.16 L_c & if\ R < R_{crit} \\ 1.373R + L_c & if\ R > R_{crit} \end{Bmatrix} \qquad (23)$$

To understand this discontinuous phenomenon, the forces acting upon the bubbles were investigated using a similar approach to Boubendir et al. [45]. They identified three lifting forces, the buoyancy force ($F_B$), pressure force ($F_P$) and momentum force ($F_M$), as well as three adhesive forces, the viscous drag force ($F_D$), added mass force ($F_{AM}$) and surface tension force ($F_S$). Due to the relatively slow growth of the bubbles in this study, the momentum force, viscous drag force and added mass force could be neglected. The buoyancy force was calculated as [46]:

$$F_B = (\rho_l - \rho_v)gV \qquad (24)$$

where *V* is volume of the bubble. The pressure force is [45]:

$$F_P = (P_v - P_l)\pi R_{base}^2 \tag{25}$$

with *P* being the pressure and $R_{base}$ being the base radius (assuming a circular geometry) of the bubble or the radius of the dry spot. The surface tension force is [47]:

$$F_S = 2\pi R_{base}\sigma \sin(\theta) \tag{26}$$

with *θ* being the contact angle. Both the pressure and surface tension forces are dependent on the radius of the base of the bubble.

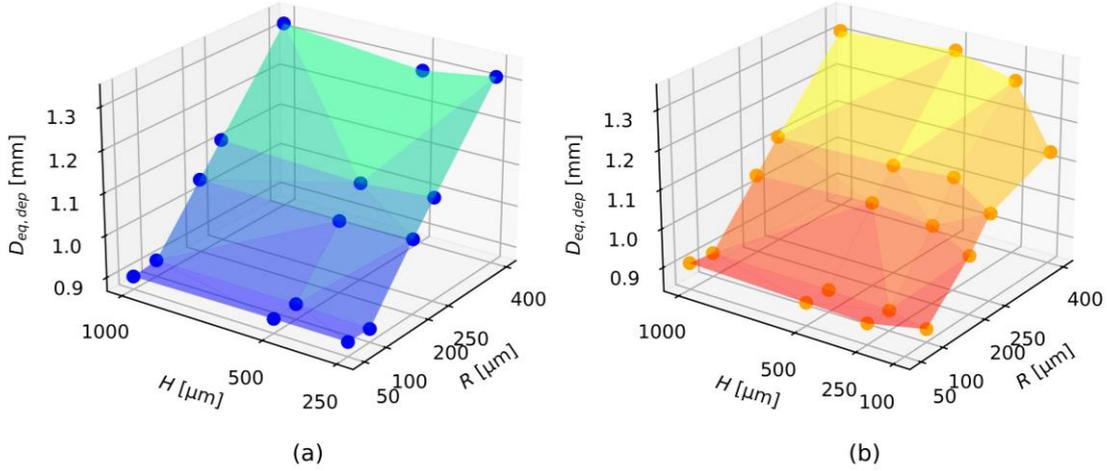

*Figure 8: Scatter plot of the equivalent bubble diameter just prior to departure for (a) cylindrical and (b) conical cavities of various cavity depths and radii.*

Figure 9 compares the variation of the base radius during a bubble's lifecycle for selected cylindrical cavities. The base radius was taken as the maximum *x*-value where the liquid–vapour interface made contact with the solid surface. As schematically indicated in (a) and (b) in Figure 9, this resulted in a large jump in the base radius curve for wide cylindrical cavities when the interface met the cavity wall, instantaneously increasing from the bubble's base to the cavity wall. For narrow cavities, after the bubble emerged from the cavity, the base radius saw a gradual increase up to a peak value ($R_{peak}$ = 275 μm), followed by a rapid decrease to the cavity radius, at which point the bubble departed. For wide cavities with *R* < $R_{peak}$ (indicated by the red data), the base radius increased after emerging from the cavity until it reached $R_{peak}$ as represented in schematic (c). Thereafter, it decreased as the bubble necking process began. Once the base of the bubble returned to the cavity radius, the process was halted (shown in schematic (d)), as the sum of buoyancy and pressure forces was not sufficient to overcome the surface tension force, which was enhanced by the larger base radius. This resulted in the bubble growing larger (along with the buoyancy force), enabling detachment. This explains the plateau in the red data in Figure 9, and the disparity in behaviour of the narrow and wide cavities. For *R* > $R_{peak}$, such as the 400 μm case (indicated by the purple data), the base radius never increased beyond the radius of the cavity, implying there was no dry-out of the horizontal surface, and therefore no rewetting. A similar investigation was also conducted for conical cavities, and the results are analysed in Section 3.3.

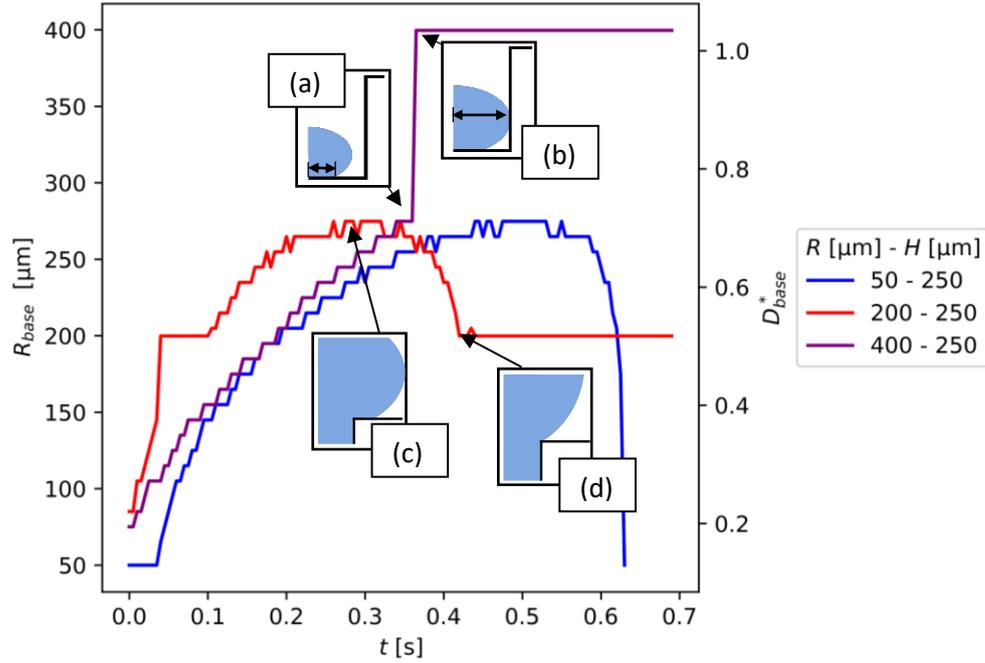

*Figure 9: Comparison of the bubble base radius during the first bubble lifecycle for cylindrical cavities with a depth of 250 μm and radii of 50 μm (blue), 200 μm (red) and 400 μm (purple). The schematics (a) and (b) indicate how the bubble base radius measurement shifts when the bubble encounters the cavity wall and (c) and (d) indicate the bubble shape at different stages of the base radius curve.*

To gain a deeper understanding of the significance of the various forces on bubble growth, the forces are compared throughout the bubble's lifecycle in Figure 10(a) for narrow and Figure 10(b) for wide cavities, where the black dotted lines represent the base radius of the bubble. It follows from Figure 10(a) that once the bubble began necking from a narrow cavity, the magnitude of the surface tension force ($F_S$) drastically decreased, and the total force ($F_T$) became positive as the bubble departed. In Figure 10(b), for a wide cavity, after necking had been halted upon reaching the cavity's edge, the surface tension force plateaued. It remained sufficiently large that the buoyancy and pressure forces could not overcome it until the bubble had reached a larger size. It should also be noted that for wide cavities, the net force never reached a substantial positive value. The reason for this is that these forces were measured at the base of the bubble.

As indicated in Figure 11(a), narrow cavities experienced necking close to the bubble base, while wide cavities (Figure 11(b)) experienced necking further above this base. This necking behaviour resulted in a variation in the amount of vapour remaining attached to the surface after departure with wider cylindrical cavities left with larger bubbles remaining after departure and vice versa for narrower cavities. Despite the variations in equivalent diameter before and after departure, the amount of vapour that departed from the cavity ($D_{dep}$) deviated little for cylindrical cavities, with $D_{dep}$ = 0.624 mm = 0.81$L_c$ (±7%) for cylindrical cavities. As discussed in Section 3.3, similar behaviour was observed for conical cavities below a certain cavity angle, however more vapour was able to detach for conical cavities with wider angles.

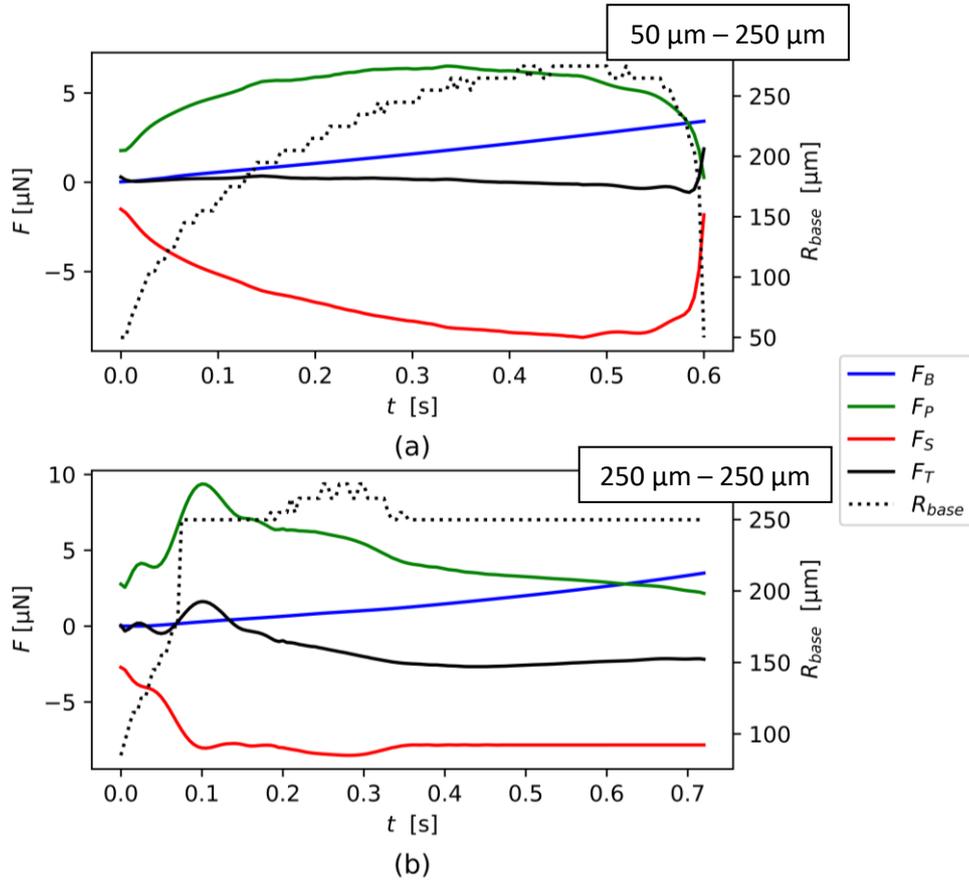

Figure 10: Representative force graphs for (a) narrow cylindrical cavities (radius: 50 μm, depth: 250 μm) and (b) wide cylindrical cavities (radius: 250 μm, depth: 250 μm). The forces due to bouyancy (blue), pressure (green) and surface tension (red) are shown including the total force (solid black) and respective bubble base radius (black dotted).

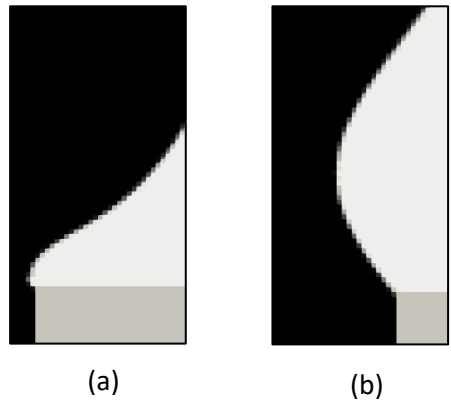

Figure 11: CFD images showing the detachment behaviour for (a) narrow cylindrical cavities and (b) wide cylindrical cavities.

The critical radius at which the bubble departure behaviour transitioned from narrow cavity behaviour ($D_{dep}$ independent of $R$) to wide cavity behaviour ($D_{dep}$ linearly dependent on $R$) was determined through a force balance:

$$F_T = (P_v - P_l)\pi R_{base}^2 - 2\pi R_{base}\sigma \sin(\theta) + (\rho_l - \rho_v)gV \qquad (27)$$

The critical radius was defined as the base radius at which the above force balance became zero. The following simplifying assumptions were made: (1) constant fluid properties ($\sigma, \rho_l, \rho_v$), (2) contact angle of 90° since this force balance focused on departure behaviour [45], (3) bubble volume ($V$) was constant and equal to the bubble volume at departure for narrow cavities ($V_{narrow}$). The reasoning behind this is that the critical radius marks the transition from narrow cavity behaviour to wide cavity behaviour and would result in a base radius value that was just able to prevent a bubble of size $V_{narrow}$ from departing. This was true when the total force was zero for a bubble of size $V_{narrow}$.

As indicated in Figure 12(a), the pressure difference ($\Delta P = P_v - P_l$) at departure decreased as the base radius at departure (equivalent to the cavity radius) increased. An exponential trendline was obtained using the average pressure difference at departure for each cavity radius (units in µm). An exponential trend was selected as it is expected that the change in pressure difference would asymptotically approach zero as the bubble radius increases, never reaching a negative pressure difference similar to the Laplace pressure.

The net force then became a function of the base radius only, as indicated in Figure 12(b). The critical radius was found to be 120 µm (approximately $L_c/6$), which corresponded to the observation that the critical radius would be between 100 µm and 200 µm. This critical radius is not expected to be universal as it is dependent on fluid properties, but it is expected that most fluids will exhibit the presence of a critical radius since it is dependent on forces commonly considered in bubble dynamics.

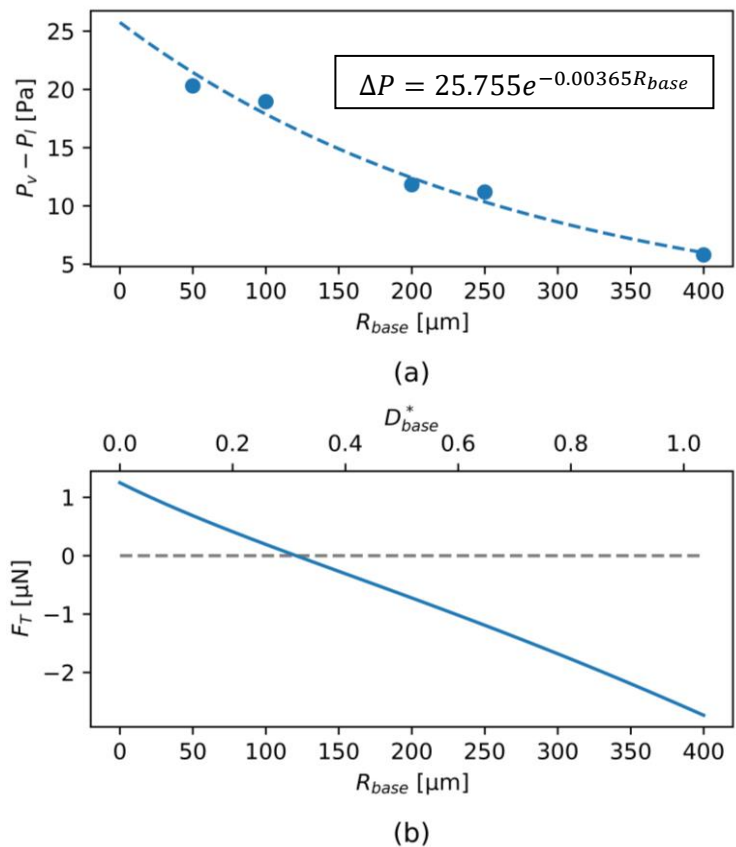

*Figure 12: Variation of (a) the pressure difference with respect to bubble base radius at departure (with exponential function fitted to data points) and (b) the total force for a given bubble base radii or non-dimensional base diameter (grey dashed line represents zero total force).*

As it was found that the cavity radius influenced the minimum bubble base radius, it is a key parameter in determining the amount of heat transferred due to surface rewetting. The amount of heat transferred to the fluid during post-departure rewetting may be formulated as [48]:

$$\dot{q}_{rw} = \int_{R}^{R_{peak}} \frac{k(T_{wall} - T_l)}{\sqrt{\pi \gamma_l}} \frac{2\pi R_{base} dR_{base}}{\sqrt{f(R_{peak}) - f(R_{base})}} \tag{28}$$

In this formulation, the function $f$, relates the time at which rewetting first occurs to the position of the interface on the surface, while variations of Equation 28 and the function $f$ are available in Demiray and Kim [48]. The amount of heat transferred during the rewetting process scales with the rewetted area ($A_{rw}$):

$$A_{rw} = \pi(R_{peak}^2 - \min(R_{base})^2) \tag{29}$$

For cylindrical cavities and conical cavities with small angles, the bubble interface did not slip down the cavity wall during the necking process, therefore the minimum $R_{base}$ value was equivalent to the cavity radius. Since the $R_{peak}$-value was observed to be fixed at 275 µm for both cylindrical and conical cavities, the rewetting area was only dependent on the cavity radius (for cylindrical and small-angled conical cavities), with narrower cavities resulting in more rewetting and cavities wider than $R_{peak}$ experiencing zero rewetting.

It follows from Equation 28 that the rewetting heat transfer is dependent on the temperature difference between the solid surface and the rewetting fluid, implying that subcooled conditions may result in higher rewetting heat transfer. Therefore, it is conjectured that systems characterised by subcooled conditions would benefit most from narrow cavities, exploiting the high rewetting area to allow for faster development of the thermal boundary layer around the bubble, promoting growth for the next lifecycle. Conversely, wider cavities may be beneficial for high-heat systems where the thermal boundary layer can quickly redevelop since, as may be inferred from Figure 11(b), these cavities result in larger bubbles remaining at the surface after detachment. With a larger initial bubble, there is a larger surface area over which heat transfer and phase change may occur and, provided there is sufficient superheated liquid around the bubble interface, which leads to more rapid growth at the early stages of the bubble lifecycle and thus a higher detachment frequency.

In terms of cavity array design, smaller cavity radii would typically be preferred as they result in smaller equivalent bubble diameters. This allows for greater nucleation site density while avoiding inter-bubble interactions which may inhibit bubble growth and nucleation [15]. Additionally, for cylindrical cavities with radii smaller than the critical radius of 120 µm, the bubble growth and departure vary negligibly allowing for an ease on manufacturing and design constraints, resulting in potential cost savings.

### 3.2 Cavity Depth

To investigate the influence of cavity depth on the bubble growth cycle, Figure 13 compares $D_{eq}$ as a function of time for different depths (represented using different line types) and radii (represented using different colours) for both cylindrical (Figure 13(a)) and conical (Figure 13(b)) cavities. A general and important trend in this figure is that the cavity depth had a negligible effect on the bubble size at departure for a given cavity radius, but increasing the cavity depth decreased the bubble growth time, as seen in the green curves in Figure 13(a) and the red curves of Figure 13(b). Figure 14(a) and Figure 14(b) compares the CFD images of the thermal boundary layer in the cavities just after seeding the bubble. A greater volume of superheated liquid existed within deeper cavities at initialisation. As the bubble grew and filled the cavity (Figure 14(c)), the surplus superheated liquid was pushed out of the cavity, bolstering the thermal boundary layer above the cavity and enabled the bubble to grow faster once it emerged from the cavity. From the red curves of Figure 13(b) it follows that the simulation results and correlation predictions diverge increasingly as the cavity angle increases. This indicates that cavity geometry needs to be considered when trying to accurately predict bubble departure from wide-angle conical cavities.

The red curves from Figure 13(a) and blue curves from Figure 13(b) sometimes deviated from this general trend, which indicated that deeper cavities can also increase the bubble departure time. These exceptions were found to be caused by the rapid bubble growth within the cavity disrupting the thermal boundary layer (as shown in Figure 14(c)), resulting in a decreased growth rate after emergence from the cavity. This is also visible from the gradient of the bubble growth curve being slightly less for the specific cases.

When comparing the predicted and numerical results, the agreement is seen to vary depending on the cavity geometry. Across all cavities the correlation of Sakashita [43] performed the best with an average deviation of 11.3% for cylindrical cavities and 15.53% for conical cavities. The correlation of Cole and Shulman [44] had the largest average deviations of 21.9% and 34.5% for cylindrical and conical cavities, respectively. Good agreement between the correlations and numerical results were observed in the supercritical radius range (a radius larger than 120 µm) and moderate depth (500 to 1000 µm). In the case of conical cavities, significant deviations were observed for large cavity angles. Although some deviations as low as 0.3% existed for some cavities, the large disparity in the deviations indicated that the correlations were not suitable for the use of cavity array boiling surfaces. For such needs, the correlations need to account for the cavity geometry, which remains as an interesting avenue for future work.

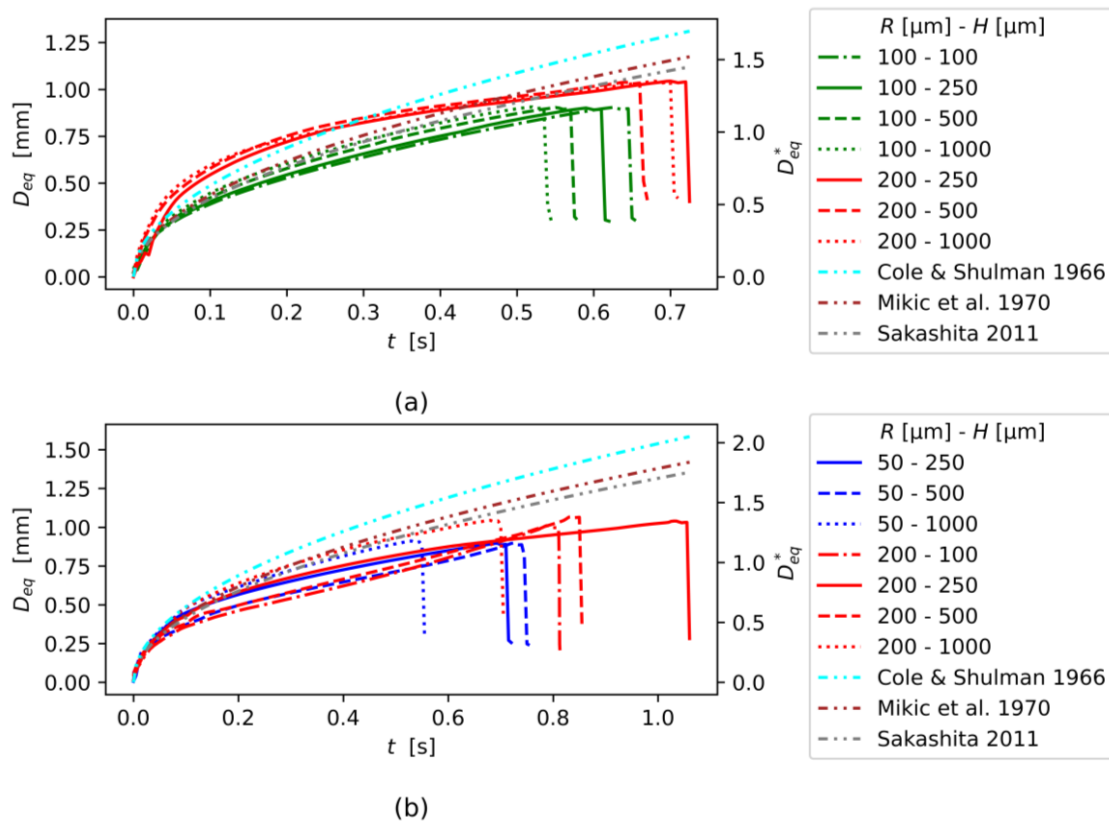

Figure 13: Bubble growth curves of the equivalent diameter highlighting the influence of (a) cylindrical and (b) conical cavity depth on bubble growth and departure for cavities with radii of 100 µm (green) and 200 µm (red). Cavity depth is indicated by line type with solid: 250 µm, dashed: 500 µm, dotted: 1000 µm and dash-dot: 100 µm.

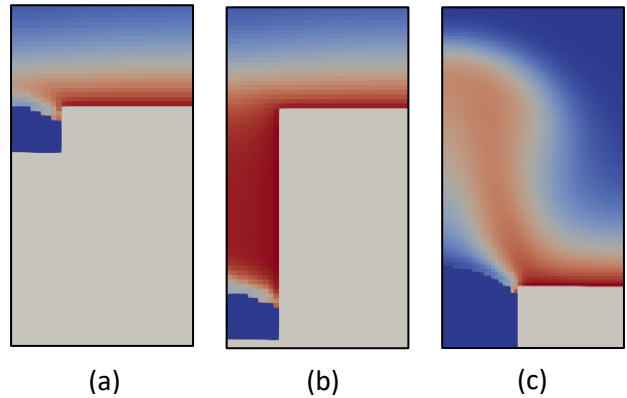

*Figure 14: CFD images highlighting how the cylindrical cavity depth and radius influences ((a) & (b)) thickness of the initial thermal boundary layer and (c) the convection of the thermal boundary layer. The cavity dimensions are: (a) radius: 100 µm, depth: 100 µm, (b) radius: 100 µm, depth: 1000 µm and (c) radius: 200 µm, depth: 1000 µm.*

It should be noted that the relationship between cavity depth and departure time was only investigated for the first bubble lifecycle. As this relationship is dependent on the bubble growth within the cavity, it is an exclusively transient behaviour. Many engineering applications operate under quasi-steady state conditions and the growth within the cavity is seldom relevant, therefore this relationship may not be of much significance to the boiling system designer. Furthermore, for quasi-steady state design, the cavity depth was found to be of potentially little influence depending on the surface superheat. Narayan et al. [13] showed that for low superheats ($Ja = 24$, which includes the liquid-vapour density ratio), there was a negligible difference in the departure frequencies from shallow (250 µm) and deep (1000 µm) cavities. However, as the superheat increased, the amount of deviation in departure frequency increased for the shallow cavity. The deeper cavity showed lower deviation causing a growing disparity in departure frequency from the two cavities.

Therefore, for the design of low superheat systems, the cylindrical cavity depth may be largely ignored. For high superheat systems, deeper cavities may offer higher stability and more regular bubble departure. As will be shown in Section 3.3, the cavity angle of conical cavities affects the departure time and cavity stability. Therefore, more attention should be given to the cavity depth when designing a system with conical cavities since the cavity depth and cavity angle are related.

## 3.3 Cavity Angle

For conical cavities, the cavity angle was influenced by changing either the radius or depth of the cavity and therefore added an additional parameter that affects the bubble dynamics of nucleate pool boiling. It was concluded from Figure 9, that for cylindrical cavities, the minimum bubble base radius was equal to the cavity radius, implying the bubble interface never traverses over the edge of the cavity. However, when comparing the bubble base radius curves for different conical cavities in Figure 15, a different trend was observed. Unlike for cylindrical cavities, the base radius curves for wide conical cavities (purple curve) did not necessarily end in a plateau. When comparing the yellow curves of Figure 15, it follows that, depending on the cavity depth and resulting cavity angle (represented using solid and dotted line types), the curve either plateaus at the cavity radius or plummets at the end of the bubble lifecycle. This variation in behaviour was observed to be dependent on cavity angle, occurring regardless of cavity radius. For a cylindrical cavity, the bubble interface undergoes a significant distortion, scaling with $\varphi$ from Equation 3, to traverse down the cavity wall and maintain the prescribed contact angle of 54°, while the surface tension of the interface resists this distortion. For a conical cavity with an angle of $\theta_{cav}$, the bubble interface undergoes a smaller distortion since $\varphi$ decreases with increasing $\theta_{cav}$ and therefore experiences less resistance from the surface tension allowing the interface to traverse down the cavity wall of wide-angle cavities during bubble departure. This implies that the minimum bubble base radius may be smaller than the cavity radius for conical cavities with large angles.

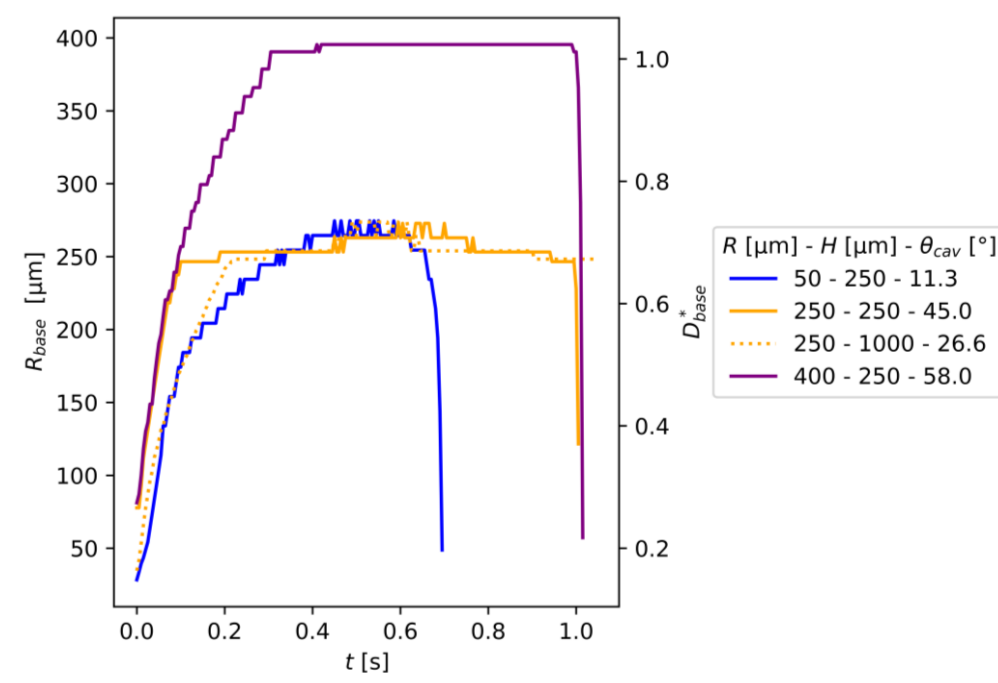

*Figure 15: Comparison of the bubble base radius during the first bubble lifecycle for concial cavities with depths of 250 µm (solid line) and 1000 µm (dashed line) and radii of 50 µm (blue), 250 µm (orange) and 400 µm (purple).*

The minimum base radius value reached during the necking process gives an indication of how far down the cavity wall the interface was able to travel before departure. As a wide array of cavity radii were tested, the minimum base radius for a given cavity was divided by the respective cavity's radius to normalise the results. For the normalised minimum base radius, a value of unity represents a bubble interface that remained pinned to the cavity edge and values lower than unity represent a bubble interface that travelled along the cavity wall as the bubble departed. Figure 16 shows the relationship between the cavity angle and this normalised minimum base radius, with larger cavity angles leading to smaller minimum base radii. Larger cavity radii also resulted in more relative slippage for any given cavity angle, with cavity angles larger than 63° leading to complete vapour departure.

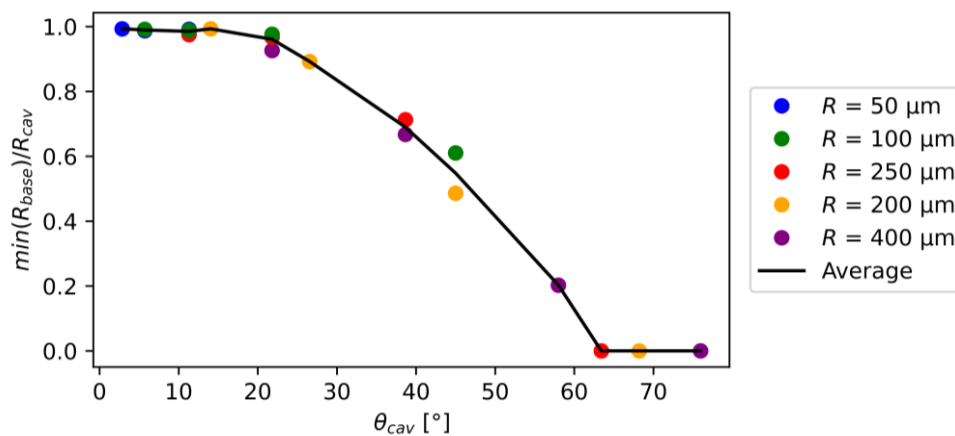

*Figure 16: Normalised minimum base radius as a function of cavity angle for conical cavities with radii of 50 µm (blue), 100 µm (green), 200 µm (red), 250 µm (orange) and 400 µm (purple).*

The ability of the bubble to traverse down the cavity wall signifies that the bubble is at risk of completely detaching, incurring a costly waiting period before the next bubble nucleation. These findings are supported by the work of Shoji and Takagi [12] who showed, using water, that conical cavities suffer from instability and frequently suffer from complete vapour detachment, with the inactivity frequency being higher for the cavity with the wider angle. Bankoff [6] and Lorenz [7] both suggested that for conical cavities to consistently retain vapour, the cavity angle (as indicted in Figure 1(b)) should be less than half the contact angle ($\theta_{cav} < \theta/2$), meaning that for R1234yf the cavity angle should be 27° to ensure vapour retention. Figure 16 shows that for cavity angles smaller than 14° the bubble interface remained pinned at the cavity edge, suggesting that for cavity angles smaller than approximately a quarter of the contact angle ($\theta_{cav} < \theta/4$), conical cavities have equivalent vapour retention potential to cylindrical cavities and so may be treated similarly to cylindrical cavities in the design process.

Unlike with cylindrical cavities that exhibited a low variation in bubble departure diameter, Figure 17 shows that the bubble departure diameter increased with an increase in cavity radius and angle. The reasoning behind this is that as the interface slipped and the surface tension force decreased, more vapour was able to join the departing bubble, resulting in an increase in the bubble departure diameter ($D_{dep}$) and a decrease in the amount of vapour left behind ($D_{rem}$). Figure 18(a) gives a visual example of the interface slipping along the cavity wall just before departure while Figure 18(b) shows the partially filled cavity after departure. Comparing Figure 11 and Figure 18 highlights the contrasting behaviour between bubble detachment from cylindrical cavities and conical cavities with interface slipping.

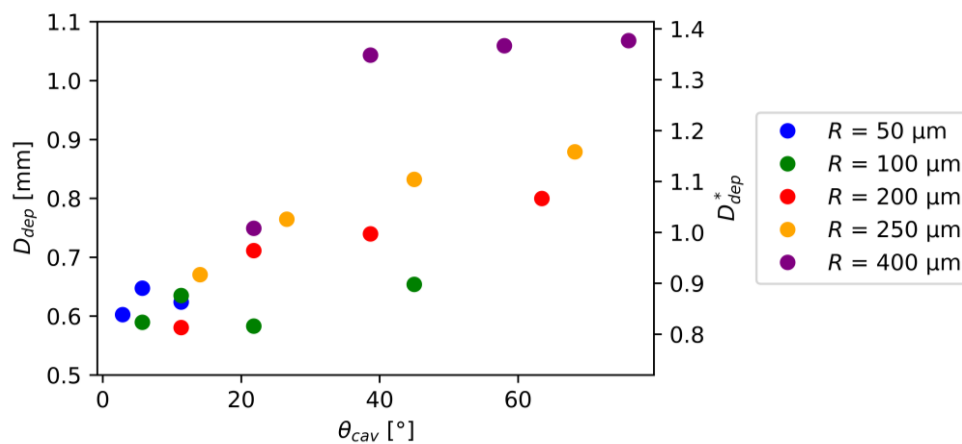

*Figure 17: Bubble departure diameter as a function of cavity angle for conical cavities with radii of 50 µm (blue), 100 µm (green), 200 µm (red), 250 µm (orange), and 400 µm (purple).*

Figure 19 compares the departure diameter of bubbles from both (a) cylindrical and (b) conical cavities for different cavity radii and depths. This figure highlights how more vapour can leave the system per bubble departure from conical cavities. This is beneficial to the overall system heat transfer as it results in more surface rewetting and more heat removed by phase change per bubble lifecycle. However, as shown in Figure 20, a larger departure diameter can also result in less vapour remaining attached to the surface and thus a larger risk of complete vapour detachment. If vapour remains attached to the surface, there is no waiting time ($t_w$) between bubble departure and the nucleation of the subsequent bubble, allowing for higher bubble departure frequencies and heat transfer rates. If all the vapour detaches from the surface, the thermal boundary layer must develop and a new bubble must nucleate, which increases the waiting time. Due to the absence of bubble growth during this waiting time, the heat transfer rate decreases and the possibility of overheating the surface, and damage to components in the case of electronic immersion cooling, increases. Therefore, minimising waiting times is important in ensuring the longevity of a system and should be taken into consideration for cavity array designs.

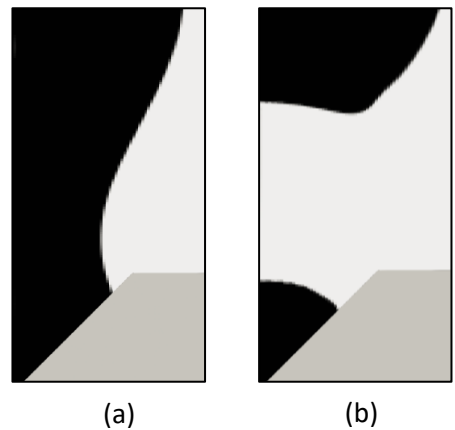

*Figure 18: CFD images showing the bubble departure from a wide-angle conical cavity with (a) just before departure and (b) just after departure.*

Figure 19 compares the departure diameter of bubbles from both (a) cylindrical and (b) conical cavities for different cavity radii and depths. This figure highlights how more vapour can leave the system per bubble departure from conical cavities. This is beneficial to the overall system heat transfer as it results in more surface rewetting and more heat removed by phase change per bubble lifecycle. However, as shown in Figure 20, a larger departure diameter can also result in less vapour remaining attached to the surface and thus a larger risk of complete vapour detachment. If vapour remains attached to the surface, there is no waiting time ($t_w$) between bubble departure and the nucleation of the subsequent bubble, allowing for higher bubble departure frequencies and heat transfer rates. If all the vapour detaches from the surface, the thermal boundary layer must develop and a new bubble must nucleate, which increases the waiting time. Due to the absence of bubble growth during this waiting time, the heat transfer rate decreases and the possibility of overheating the surface, and damage to components in the case of electronic immersion cooling, increases. Therefore, minimising waiting times is important in ensuring the longevity of a system and should be taken into consideration for cavity array designs.

While larger cavity angles may result in more surface rewetting and larger bubble departure diameters, the Bankoff [6] criterion is recommended for the design of conical cavity arrays to ensure cavity stability. For more conservative approach, the designer could ensure a cavity angle below 14° (for R1234yf) as this resulted in stability equivalent to a cylindrical cavity. Furthermore, this allows for a simpler design process as the rewetting area may be calculated using the cavity radius only.

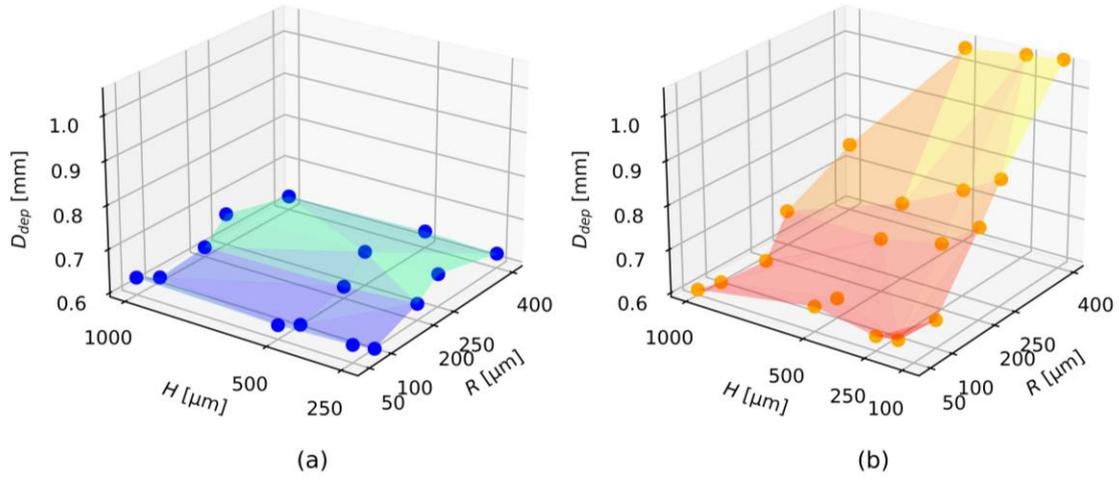

Figure 19: Scatter plot of the bubble departure diameter for (a) cylindrical and (b) conical cavities of various cavity depths and radii.

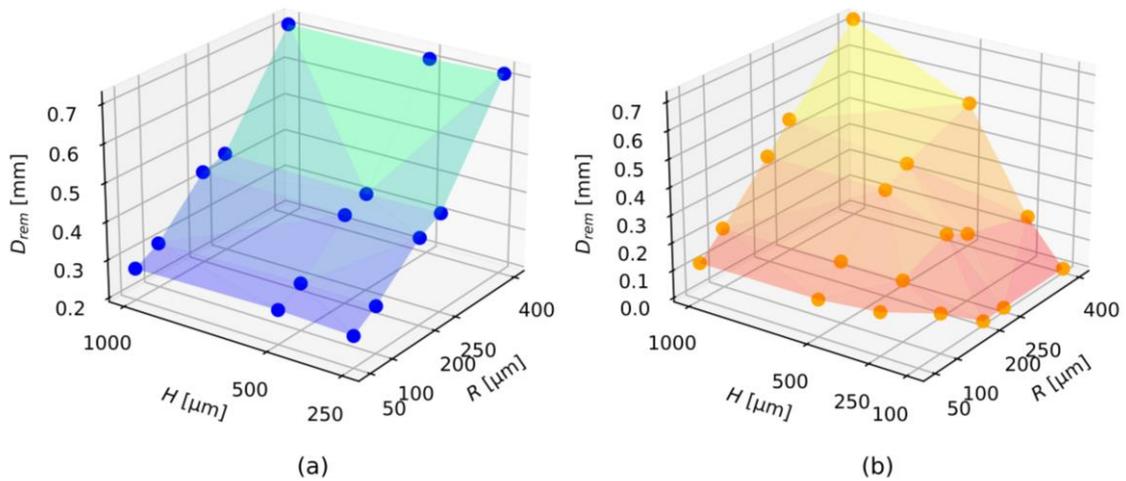

Figure 20: Scatter plot of the equivalent diameter of the bubble remaining attached to the surface after departure for (a) cylindrical and (b) conical cavities of various cavity depths and radii.

# 4 Conclusions

This study numerically investigated the effect of cylindrical and conical cavity geometries on the bubble dynamics associated with the nucleate pool boiling of R1234yf. At a fixed input heat flux of 28 kW/m$^2$, the cavity radius was varied between 50 µm and 400 µm and the cavity depth between 100 µm and 1000 µm. The cavity angle of the conical cavities resulted from the choice in cavity radius and depth. The following conclusions were made from the single bubble dynamics:

- The equivalent diameter at departure did not vary for cavities with radii below a critical radius of 120 µm (approximately one sixth of the capillary length). However, the equivalent diameter at departure increased linearly with increasing cavity radius above this critical radius. Due to the necking process before bubble departure, the bubble base receded along the surface, reaching a minimum radius equal to the respective cavity radius for cylindrical cavities. Therefore, the area of surface rewetting was a function of cavity radius only. Smaller cavity radii enabled more rewetting, while a radius larger than the peak bubble base radius of 275 µm experienced no rewetting during bubble detachment.

- The cavity geometry led to variations in bubble growth time due to differences in the initial thermal boundary layer thickness. This variation is assumed to become negligible for systems with low Jakob numbers at quasi-steady state conditions.
- The cavity angle of conical cavities was found to have a negligible effect on the bubble departure diameter for angles less than a quarter of the contact angle. These conical cavities also had similar vapour retention potential to cylindrical cavities. As the angle was increased further, the bubble departure diameter increased, with the amount of vapour remaining in the cavity after departure decreasing. For cavity angles large than 60°, complete vapour departure was observed.

When translating the fundamental knowledge obtained for single cavities to cavity array design, the following recommendations were made:

- Cylindrical cavities are recommended for more stable and predictable heat transfer rates as they have a higher likelihood of encouraging nucleation.
- Smaller cavity radii are favourable for subcooled conditions where rewetting heat transfer is most significant. Furthermore, smaller cavity radii are preferable as they allow for greater nucleation site density while resulting in the same bubble departure diameter as larger cavity radii.
- Larger cavity radii are recommended for high superheat conditions as they typically resulted in larger bubbles, with more bubble surface area remaining after detachment (dependent on cavity angle for conical cavities), which could enhance bubble growth if the thermal boundary layer was sufficiently developed.
- For cavities with radii below the critical radius of 120 µm, a negligible variation in the resulting bubble dynamics was found and so less stringent manufacturing tolerances may be utilised, potentially saving on costs.
- It was found that the cavity depth was not an important design parameter for cylindrical cavities but should be accounted for when using conical cavities as it affected the departure time as well as stability.
- Conical cavity array designs are more complicated as the larger cavity angles may allow for larger bubble departure diameters at the cost of nucleation site density as well as cavity stability, which both affects the overall heat transfer rate and longevity of the system.

# Conflict of Interest

The authors have no conflicts to disclose.

# Author Contributions

**M.S. Whiting**: Methodology, Validation, Investigation, Visualization, Data curation, Writing – original draft. **W.J. van den Bergh**: Validation, Investigation, Writing – review & editing. **P.E. Theodorakis**: Methodology, Software, Supervision, Project administration, Resources, Writing – review & editing, Funding acquisition. **M. Everts**: Conceptualization, Supervision, Project administration, Resources Writing – review & editing, Funding acquisition.

# Data availability

The data that support the findings of this study are available from the corresponding author upon reasonable request.

# Acknowledgements

This project was funded by the European Union's Horizon 2020 research and innovation programme under the Marie Skłodowska-Curie grant agreement No. 778104 (Project ThermaSMART) and was also supported by the Polish high-performance computing infrastructure PLGrid (HPC Centers: ACK Cyfronet AGH, WCSS) for providing computer facilities and support within computational grant no. PLG/2023/016280.